\documentclass[a4paper,11pt]{article}
\usepackage{pos}
\usepackage{packages}

\title{Progress on nucleon transition matrix elements with a lattice QCD variational analysis}
\author*[a]{Lorenzo Barca}
\author[b]{Gunnar S. Bali}
\author[b]{Sara Collins}
\affiliation[a]{John von Neumann-Institut f\"ur Computing NIC, 
                Deutsches Elektronen-Synchrotron DESY, Platanenallee 6, 15738 Zeuthen, Germany}
\affiliation[b]{Fakult\"at f\"ur Physik, Universit\"at Regensburg, Universit\"atsstraße 31, 93053 Regensburg, Germany}

\emailAdd{lorenzo.barca@desy.de}
\emailAdd{gunnar.bali@ur.de}
\emailAdd{sara.collins@ur.de}

\abstract{
Nucleon weak matrix elements can be extracted from nucleon correlation functions with lattice QCD simulations. 
The signal-to-noise ratio prohibits the analysis at large source-sink separations and as a consequence, 
excited state contamination affects the extraction of the nucleon matrix elements. Chiral perturbation theory (ChPT) 
suggests that the dominant contamination in some of these channels is due to $N\pi$ states where the pion carries 
the same momentum of the current. In this talk, we report updates on the variational analysis with $qqq$-operators 
(nucleon-like) and $(qqq)(\bar{q}q)$-operators (nucleon-pion-like) where we report for the first time some preliminary 
results of $\langle N\pi| \mathcal{J}| N \rangle $, modulo some kinematic and volume factors, and we compare the results 
against ChPT. This pilot study is performed on a CLS ensemble with $N_f=3$, $m_\pi \approx 420~\mathrm{MeV}$, 
$a\approx 0.1~\mathrm{fm}$ and $T=2L\approx 4.8~\mathrm{fm}$.
}

\FullConference{%
  EuroPLEx Final Conference \\
  11–15 Sept 2023\\
  Institute for Physics, Humboldt University Berlin
}

\begin{document}
\maketitle

\section{Introduction}\label{sec1}

The next-generation neutrino oscillation experiments DUNE \cite{DUNE:2015lol, DUNE:2020lwj} and Hyper-Kamiokande \cite{Hyper-Kamiokande:2022smq} 
will address fundamental questions in particle physics and cosmology,
including whether neutrinos oscillate differently from antineutrinos.
CP violation in the lepton sector \cite{Pontecorvo:1967fh} is necessary for a 
possible Standard Model explanation of the matter-antimatter asymmetry in our universe \cite{Sakharov:1967dj}.
Other prospects for physics searches at DUNE include more precise studies of the neutrino mass hierarchy, 
supernova neutrino bursts \cite{DUNE:2020zfm}, and BSM physics \cite{DUNE:2020fgq}.
Of fundamental importance are the parameters that describe the cross sections of neutrino-nucleus scattering data,
as very high precision is required in the analysis of such rare weak processes.
Nuclear models relate the neutrino-nucleus scattering cross-section, whose events occur at the detector level,
to the neutrino-nucleon scattering cross-section, which is parametrised in terms of nucleon form factors \cite{Kopp:2024yvh, Lovato:2023khk}.

MiniBooNE, a previous terrestrial neutrino oscillation experiment, observed a significant production of nucleon-pion final states (CC-1$\pi$)
in addition to quasielastic charged current processes (CCQE) \cite{PhysRevD.81.092005}.
These additional states are produced indirectly via resonance states, such as $\nu_\mu + n \to \mu^- + \Delta^+ (\to \pi^+ + n)$ 
in the intermediate energy range $E_\nu \approx 0.1 - 20~\mathrm{GeV}$ \cite{RevModPhys.84.1307}.
Achieving a few-percent accuracy in determining (anti)neutrino oscillation parameters requires precise knowledge of 
the electroweak transitions $N\to \Delta$, $N\to N^*$ and $N\to N\pi$ in the non-perturbative regime \cite{Simons:2022ltq}.

First principles calculations based on lattice QCD simulations can provide the form factors 
that enter in the cross-sections \cite{Ruso:2022qes, USQCD:2022mmc}.
Nucleon form factors are extracted from ratios of nucleon three- and two-point functions computed using Monte Carlo techniques.
A major challenge of lattice nucleon structure calculations is the problem of excited state contamination.
In principle, this can be overcome by evaluating  nucleon correlation functions at large source-sink separations, as the excited state contamination decays exponentially with the Euclidean time separation. 
Unfortunately, the signal-to-noise ratio deteriorates with the distance between nucleon operators, 
making it impractical to analyse the data at long distances.
A complete variational analysis which involves constructing the nucleon spectrum using 
the Lüscher method \cite{Luscher:1990ux, Luscher:1991cf} and then using the GEVP solutions to diagonalise 
the interpolating operators \cite{Blossier:2009kd, Bulava:2011yz}, can potentially solve this problem.
This more robust approach requires an analysis with both single- ($N$) and 
multi-hadron ($N\pi$, $N\pi\pi$, ...) operators, making lattice simulations demanding
due to the increased number of Wick contractions and Dirac propagators.

Chiral perturbation theory (ChPT) can guide us in understanding which states contribute the most to the contamination of the nucleon three-point functions and, 
consequently, which interpolating operators are most relevant for the variational analysis.
A reliable estimation of the excited state contamination and other systematic effects is essential for the accurate extraction of the $N \to N$ electroweak form factors.
At leading order in ChPT, it was found that $N\pi$ states in the kinematic configuration where the pion carries the same momentum as the current represent the dominant contribution \cite{Bar:2018xyi, Bar:2019gfx}. Other $N\pi$ states as well as $N\pi\pi$ and nucleon excitations will also be relevant to reach few percent accuracy.
Multistate fits informed by ChPT have proven to be successful in determining form factors that satisfy the Pion Pole Dominance (PPD)
and the generalised Goldberger-Treiman (also known as PCAC) relations \cite{RQCD:2019jai, Jang:2019vkm}.

In this study, we carry out a variational analysis utilising just a single $qqq$ (three-quark or $N$-like) interpolating 
operator and a single $(qqq)(\bar{q}q)$ (five-quark or $N\pi$-like) interpolating operator.
Specifically, we consider the $N\pi$-like operator where the pion carries the same momentum as the current, 
aiming to verify the ChPT prediction regarding its large contribution.
We compute nucleon matrix elements $\langle N| \mathcal{J} | N \rangle$, previously presented in \cite{Barca:2022uhi} 
and, for the first time, we present preliminary results for positive-parity transition matrix elements $\langle N\pi| \mathcal{J} | N \rangle$ in a finite volume, up to some kinematic factors. 
Here, $|N\rangle$ denotes the nucleon ground state and $|N\pi\rangle$ a physical nucleon-pion finite volume state.
This work provides an update of the results of \cite{Barca:2021iak, Barca:2022uhi}.
In sec.~\ref{sec2}, we provide an overview of the extraction of nucleon form factors from nucleon matrix elements using the standard ratio method,
along with a discussion of key results from ChPT in these channels.
In sec.~\ref{sec3}, we discuss how we use the variational analysis to determine the matrix elements by projecting the operators to the desired eigenstate. 
In sec.~\ref{sec4}, we present preliminary results for $\langle N\pi | \mathcal{J} | N\rangle $ at zero spatial momentum transfer.
Finally, in sec.~\ref{sec5}, we draw conclusions and discuss key observations of this analysis.

\section{Nucleon matrix elements}\label{sec2}
\subsection{Ratio method}
The nucleon matrix elements $\langle N(\pp') | \mathcal{J}(\qq) | N(\pp) \rangle $ 
appear in the spectral decomposition of the nucleon three-point correlation functions
\begin{equation}\label{nucleon_3pt}
C^{k, \mathcal{J}}_{3pt}(\pp', t; \qq, \tau) ~=~ \mathbb{P}^{+}_k  ~ \langle \mathrm{O}_{3q}(\pp', t) ~\mathcal{J}(\qq, \tau)~ \bar{\mathrm{O}}_{3q}(\pp, 0) \rangle~,
\end{equation}
as it can be seen by inserting two complete sets of states.
In this expression, $\mathrm{O}_{3q}$ is a $3$-quark operator that has the nucleon quantum numbers $J^{PC}= \frac{1^{}}{2}^{++}$ and $\mathcal{J}$ is the local current projected to momentum $\qq=\pp'-\pp$.
The matrix $\mathbb{P}^+_k$ projects the operators to positive parity and it aligns the spin along the direction $k$. 
In eq.~\eqref{nucleon_3pt}, $t$ is the source-sink separation between the two nucleon operators and $\tau$ is the intermediate time where the current is located.
Additionally, one computes the nucleon two-point functions
\begin{equation}\label{nucleon_2pt}
C_{2pt}(\pp, t) ~=
~\mathbb{P}^{+} ~ \langle \mathrm{O}_{3q}(\pp, t)~\bar{\mathrm{O}}_{3q}(\pp, 0) \rangle~,
\end{equation}
where $\mathbb{P}^{+} $ is a positive-parity projector and
the spectral decomposition reads explicitly
\begin{equation}
\label{spectral_decomposition_2pt}
C_{2pt}(\pp, t) ~=~
\frac{|\langle \Omega | \mathrm{O}^+_{3q} | N(\pp) \rangle|^2}{2E_N V}
e^{-E_Nt}
~+~
\sum_\kk
\frac{|\langle \Omega | \mathrm{O}^+_{3q} | N(\pp+\kk)\pi(-\kk) \rangle|^2}{(2E_N V) (2E_\pi V)}
e^{-E_{N\pi }t}
~+~ \dots
\end{equation}
where $\oo^+_{3q}=\mathbb{P}^{+} \oo_{3q}$ and $|\Omega\rangle$ is the vacuum state.
In this expression, to simplify the notation we include only the ground state $N$ and the tower of $N\pi$ states with total momentum $\pp$, and in the ellipses we include all the other states in the spectrum.
Using nucleon two- and three-point functions, we construct the following ratio
\begin{equation}\label{ratio_method}
R^{k, \mathcal{J}}(\pp', t; \qq, \tau)
:=  ~
\frac{C_{3pt}^{k, \mathcal{J}}(\pp', t; \qq, \tau)}{C_{2pt}(\pp', t)} \
\sqrt{\frac{C_{2pt}(\pp', \tau)~C_{2pt}(\pp', t)~C_{2pt}(\pp, t-\tau)}
{C_{2pt}(\pp, \tau) ~C_{2pt}(\pp, t) ~C_{2pt}(\pp', t-\tau)}}~.
\end{equation}
By employing the spectral decomposition of the nucleon three- and two-point functions, it is straightfordward to show that the nucleon ground state contribution to eq.~\eqref{ratio_method} is  independent of $t$ and $\tau$, and  proportional to $\langle N(\pp') | \mathcal{J}(\qq) | N(\pp) \rangle$. The latter can be decomposed in terms of nucleon form factors through Lorentz decomposition, which depends on the intermediate current. For pseudoscalar and axial-vector currents, the Lorentz decompositions in Euclidean space-time are
\begin{align}
	\label{Lorentz_decomposition_pseudoscalar}
	&\langle N(\pp')| {\mathcal{P}(\qq) } | N(\pp) \rangle=
	\bar{u}_{N}(\pp')~\left[\gamma_5 G_P(Q^2)\right] ~u_{N}(\pp)~,
	\\
	\label{Lorentz_decomposition_axial}
	&\langle N(\pp')| {\mathcal{A}_\mu(\qq) } | N(\pp) \rangle=
	\bar{u}_{N}(\pp')~
	\left[
	\gamma_\mu \gamma_5 G_A(Q^2) + \frac{q_\mu}{2m_N} \gamma_5 \widetilde{G}_P(Q^2) \right]
	~u_{N}(\pp)~,
\end{align}
where $q_\mu=p'_\mu-p_\mu$ and $Q^2=-q^\mu q_\mu$ is the four-momentum transfer.
In eq.~\eqref{Lorentz_decomposition_pseudoscalar}, $G_P$ is the pseudoscalar form factor, 
while in eq.~\eqref{Lorentz_decomposition_axial},
$G_A$ and $\widetilde{G}_P$ are the axial and induced pseudoscalar form factors, respectively.
If we write the Lorentz decomposition like
$\langle N(\pp')| {\mathcal{J}(\qq) } | N(\pp) \rangle = \bar{u}_N(\pp') \left[FF_{N\mathcal{J}N}\right] u_N(\pp)$, the spectral decomposition of the ratio in eq.~\eqref{ratio_method} yields
\begin{equation}\label{spectral_ratio_method_nn}
R^{k, \mathcal{J}}(\pp', t; \qq, \tau)
=
\sqrt{\frac{(E'_NV)(E_NV)}{(E'_N+m_N)(E_N+m_N)}}
\frac{\mathrm{Tr}\left\{\mathbb{P}^+_k (\slashed{p}'+m_N)
\left[FF_{N\mathcal{J}N}\right]
(\slashed{p}+m_N) \right\}}{(2E'_NV)(2E_NV)}
~+~ \dots,
\end{equation}
where we are considering only the nucleon ground state, while the contributions from excited states, which decay exponentially with time, are included in the ellipses. 
The nucleon form factors in $\left[FF_{N\mathcal{J}N}\right]$ can be extracted from the ratio in eq.~\eqref{ratio_method} at large $t$ and $\tau$, as the contribution from excited states is reduced.
However, the correlation functions in eqs.~\eqref{nucleon_3pt}--\eqref{nucleon_2pt} are estimated with Monte Carlo simulations and the error associated with the measurement decreases more slowly than the signal. Thus, the degradation of the signal-to-noise ratio prohibits the analysis at large source-sink separations. In principle, the error can be systematically reduced as it scales with $\mathcal{O}(1/\sqrt{N})$, where $N$ is the number of gauge configurations or statistical samples. However, it is impractical with the current computers and algorithms to compute nucleon correlation functions with $N \gg \mathcal{O}(10^4)$, and thus the statistical sample size is usually limited to $10^3 \lesssim N \lesssim 10^4$.
Due to this constraint, the extraction of nucleon form factors from the ratio in eq.~\eqref{ratio_method} is affected by excited state contamination (ESC). To resolve the nucleon ground state, one needs a robust method that takes into account all the relevant excited states. 

\subsection{$N\pi$ states contamination in nucleon correlation functions from ChPT}
ChPT can provide guidance on the contribution of specific states to the correlation functions. Since ChPT is an expansion for small pion masses and momenta, its predictions are most reliable towards physical pion masses and small momenta.
In \cite{Bar:2015zwa}, it was estimated with leading order in ChPT (LO-ChPT) that the contamination of the tower of $N(\pp)\pi(-\pp)$ states to the zero-momentum nucleon two-point function is at the level of a few percent for close to physical pion masses ($\approx 200~\mathrm{MeV}$) and source-sink separations of $t \approx 0.5~\mathrm{fm}$.
The contribution of $N\pi\pi$ states to the nucleon two-point functions is estimated to be even smaller at the per mille level \cite{Bar:2018wco}. Multiparticle states are usually suppressed by a volume factor for each additional state.
While LO-ChPT predicts that $N\pi$ states do not contribute much to the nucleon two-point functions at sufficently large distance, their contribution to the nucleon three-point functions can be very large.
In \cite{Bar:2018xyi, Bar:2019gfx}, it was predicted with LO-ChPT that the $N\pi$ states contribution depends on the momentum transfer $Q^2$ and it can rise up to $40\%$ for physical pion masses at $t=2~\mathrm{fm}$. Most of this contribution is due to $N\pi$ states which are either created at the source or at the sink and where the pion has the same momentum as the current. 
These current-enhanced diagrams are not volume suppressed. 
More explicitly, in lattice simulations the nucleon operators at the sink are usually projected to $\pp'=\zzero$, and thus the momentum of the source operator is $\pp=-\qq$ by momentum conservation. With these kinematics, the dominant $N\pi$ states at the sink have total momentum $\pp'=\zzero$ and are in the momentum configuration $N(-\qq)\pi(\qq)$, while at the source they are in the momentum configuration $N(\zzero)\pi(-\qq)$.
In the forward limit ($\qq=\zzero$), ChPT predicts still a large $N\pi$ states contamination in specific channels when the sink and source nucleon operators have non-zero momentum $\pp'=\pp \neq \zzero$, see \cite{RQCD:2019jai}.
This was confirmed numerically in \cite{Barca:2022uhi}.
The remaining tower of interacting $N\pi$ states with other relative momenta and $N(\zzero)\pi(\zzero)\pi(\zzero)$ states can further contribute to contaminate the nucleon three-point correlation functions up to few percent at the physical point, and they must be taken into account for high-precision determination of the nucleon form factors.
Resonance contributions in a finite volume like the Roper in the positive-parity nucleon channel must also be considered and effective field theories provide guidance on the interacting energy levels \cite{Severt:2022jtg}.

\section{Variational analysis}\label{sec3}
The variational analysis consists of constructing a basis of $n$ operators $\mathrm{O}_i$ that have the quantum numbers of the nucleon and which are used to compute a matrix of two-point correlation functions
\begin{equation}
\mathbb{C}_{2pt}(\pp, t)_{ij} = 
~\mathbb{P}^{+} ~ \langle \mathrm{O}_i(\pp, t)~\bar{\mathrm{O}}_j(\pp, 0) \rangle~.
\end{equation}
This matrix is then employed to solve the Generalised EigenValue Problem (GEVP)
\begin{equation}\label{gevp}
\mathbb{C}_{2pt}(\pp, t) V(\pp, t; t_0) = \mathbb{C}_{2pt}(\pp, t_0) \Lambda(\pp, t; t_0) V(\pp, t; t_0)~
\end{equation}
to find the matrices of eigenvalues $\Lambda(\pp, t; t_0)=\mathrm{diag}\left(\lambda^1(\pp, t;t_0), ..., \lambda^n(\pp, t;t_0)\right)$ and eigenvectors $V(\pp, t; t_0)=\left(\mathbf{v}^1(\pp, t; t_0), \dots, \mathbf{v}^n(\pp, t; t_0) \right)$. Each eigenvalue decays exponentially with the energy of the specific state in the spectrum $\lambda^{\alpha} \propto e^{-E_\alpha (t-t_0)}$. The $n$-dimensional eigenvectors are orthogonal with respect to the scalar product
\begin{equation}\label{normalised_eigenvectors}
\left(\tilde{v}(\pp, t; t_0), \mathbb{C}_{2pt}(\pp, t_0) \tilde{v}(\pp, t; t_0) \right) 
=
\sum_{i, j}^{n} \tilde{v}_i^{*\alpha}(\pp, t; t_0) ~\mathbb{C}_{2pt}(\pp, t_0)_{ij} ~ \tilde{v}_j^{\beta}(\pp, t; t_0) = \delta^{\alpha \beta}~,
\end{equation}
where we have normalized $\tilde{v}^\alpha = 
\left(v^\alpha, \mathbb{C}_{2pt} v^\alpha\right)^{-1/2} v^\alpha$.
Since our basis is finite and cannot span the infinite dimensional Hilbert space, the $\alpha$-th eigensolutions are correct up to $\mathcal{O}(e^{-(E_{n+1} - E_\alpha)t})$, where $(n+1)$ labels the first state that is missing in our finite basis. 
The GEVP eigenvectors can be used to diagonalise the correlation matrix and to construct an operator that has overlap predominantly with a desired state $\alpha$ in the spectrum.
To construct the GEVP-improved operator the following linear combination of eigenvectors and operators must be considered:
\begin{equation}\label{gevp_operator}
\mathrm{O}^{\alpha}(\pp, t)
=
\sum_i^n v_i^{\alpha}(\pp, t_\alpha; t_0)\mathrm{O}_i(\pp, t)~.
\end{equation}
With a sufficiently large (and orthogonal) basis the system will be completely diagonalised and 
the eigenvectors will be time independent already from $t\approx 0.0~\mathrm{fm}$.
However, due to the finite size of the operator basis, the eigenvectors are usually not time independent as remaining excitations are not completely removed at small distances. 
The linear combination is then constructed using eigenvectors at a reference time $t_\alpha$ such that a plateau in the effective masses of the eigenvalues sets in with the energy of the state $\alpha$. 
After the eigenstate projection in eq.~\eqref{gevp_operator}, we expect that
\begin{equation}
\bar{\mathrm{O}}^\alpha | \Omega \rangle \approx Z^\alpha |\alpha\rangle~,
\end{equation}
where the approximation is due to the finite size of our operator basis.
In particular, the GEVP-improved operators are then used to compute GEVP-improved two-point functions, which read
\begin{equation}
C_{2pt}(\pp, t)^{\alpha}\label{gevp_corr2pt} ~=
~ \langle \mathrm{O}^\alpha(\pp, t) ~ \bar{\mathrm{O}}^\alpha(\pp, 0) \rangle
= \sum_{i, j}
~ v_i^{\alpha}(\pp, t_\alpha ;t_0) \langle \mathrm{O}_i(\pp, t) ~ \bar{\mathrm{O}}_j(\pp, 0)\rangle ~ v_j^{\alpha}(\pp, t_\alpha;t_0)~.
\end{equation}
Similarly, one can adopt the eigenstate-projected operators to compute the GEVP-improved three-point functions
\begin{equation}\label{gevp_corr3pt}
C^{k, \mathcal{J}}_{3pt}(\pp', t; \qq, \tau)^{\alpha \beta} ~=
\mathbb{P}^+_k
~ \langle \mathrm{O}^\alpha(\pp', t)~\mathcal{J}(\qq, \tau) ~ \bar{\mathrm{O}}^\beta(\pp, 0)\rangle ~.
\end{equation}
In order for the GEVP to resolve a finite number of $n$ states in the nucleon spectrum, the operator basis must be as orthogonal as possible, i.e., the operators must be constructed to have major overlap with different states in the nucleon spectrum.
To extract nucleon matrix elements $\langle N'|\mathcal{J}|N\rangle $ with the variational method, we project the correlation functions to the nucleon eigenstate, i.e. $\alpha=\beta=N$ in eqs.~\eqref{gevp_corr2pt}-\eqref{gevp_corr3pt}, and we 
construct the ratio
\begin{equation}\label{gevp_ratio_nn}
R^{k, \mathcal{J}}(\pp', t; \qq, \tau)^{{NN}}
=  ~
\frac{C_{3pt}^{j, \mathcal{J}}(\pp', t; \qq, \tau)^{{N N}}}{C_{2pt}(\pp', t)^{N}} \
\sqrt{\frac{C_{2pt}(\pp', \tau)^{N}~C_{2pt}(\pp', t)^{N}~C_{2pt}(\pp, t-\tau)^{N}}
{C_{2pt}(\pp, \tau)^{N} ~C_{2pt}(\pp, t)^{N} ~C_{2pt}(\pp', t-\tau)^{N}}}~.
\end{equation}
This expression is the same as in eq.~\eqref{ratio_method}, except for the fact that the correlation functions are now projected to the nucleon state ($N$).
Therefore, the exponential corrections from higher energy states included in the ellipses of eq.~\eqref{spectral_ratio_method_nn} are removed, 
provided that the basis is good enough to resolve the $n$ states in the spectrum.
Similarly, to extract transition matrix elements $\langle N\pi | \mathcal{J} | N \rangle$, we project the correlation functions both to $N$ and $N\pi$ and construct the ratio
\begin{equation}\label{gevp_ratio_npin}
R^{k, \mathcal{J}}(\pp', t; \qq, \tau)^{{N\pi N}}
=  ~
\frac{C_{3pt}^{j, \mathcal{J}}(\pp', t; \qq, \tau)^{{N\pi N}}}{C_{2pt}(\pp', t)^{N\pi}} \
\sqrt{\frac{C_{2pt}(\pp', \tau)^{N\pi}~C_{2pt}(\pp', t)^{N\pi}~C_{2pt}(\pp, t-\tau)^{N}}
{C_{2pt}(\pp, \tau)^{N} ~C_{2pt}(\pp, t)^{N} ~C_{2pt}(\pp', t-\tau)^{N\pi}}}~.
\end{equation}
A similar ratio without GEVP-improved correlators and with $\pp'=\pp=\qq=\zzero$ was employed in \cite{Wang:2023omf} to determine the $N\pi$ contribution in electric polarizabilities ($N \gamma^*\to N\pi$).
The three-point function in this expression is projected at the source to the $N$ eigenstate and at the sink to the $N\pi$ eigenstate. By applying the spectral decomposition to the correlation functions in eq.~\eqref{gevp_ratio_npin} with the normalisation convention defined in eq.~\eqref{spectral_decomposition_2pt}, we obtain the expression
\begin{align}
\nonumber
R^{k, \mathcal{J}}(\pp', t; \qq, \tau)^{{N\pi N}}
= &~
\sqrt{\frac{(2E'_NV)(2E'_\pi V)(2E_NV)}{2(E_N+m_N)
\sum_{\sigma} u^\sigma_{N\pi}(\pp') \bar{u}^{\sigma}_{N\pi}(\pp') }}
~\times~
\\
\times&~
\frac{\mathrm{Tr}\left[\mathbb{P}^+_k
\sum_\sigma u^\sigma_{N\pi}(\pp') \bar{u}^{\sigma}_{N\pi}(\pp')
\left[FF_{N\pi \mathcal{J}N} \right] 
(\slashed{p}+m_N) \right]}{(2E'_NV)(2E_NV)}~
\end{align}
where $\left[FF_{N\pi \mathcal{J}N} \right] $ is the decomposition of $\langle N\pi| \mathcal{J} | N\rangle $ in terms of form factors as in eqs.~\eqref{Lorentz_decomposition_pseudoscalar}-\eqref{Lorentz_decomposition_axial},
\begin{equation}
\langle N\pi| \mathcal{J} | N\rangle = u_{N\pi} \left[FF_{N\pi \mathcal{J}N} \right]  u_N
\end{equation}
In principle, one should include all the relevant operators to reconstruct the nucleon spectrum ($N$, $N^*$, $N\pi$, $N\pi\pi$, $\Delta \pi$, ...) and use the Lüscher method \cite{Luscher:1986pf} to compute the phase shift of $N\pi$ scattering in the Roper channel from the energy levels in a finite volume.
In particular, multi-hadron operators must be constructed with relative momenta up to some value, above which their effect is expected to be negligible.
This would give more confidence that no relevant operator is missing in the variational basis and that the eigenvectors and eigenvalues are correct up to negligible exponential corrections. Only then, the GEVP eigenvectors can be used to construct GEVP-projected two-point and three-point functions.
An example of systematic effects in the reconstruction of the spectra due to missing relevant operators in the GEVP basis can be seen for example in Fig.~6 of \cite{Wilson:2015dqa} in the $\rho$-resonance channel.

While this approach of reconstructing the spectrum is very sound, current algorithms make it impractical to realise.
The difficulty lies in the computation of many all-to-all Dirac propagators that are needed to evaluate two-point and three-point functions for two- and three-particle operators. 
In addition, the computational expense of computing a symmetric matrix of two-point or three-point functions increases with $n\times (n+1)/2$, with $n$ being the number of operators in the basis. 
Therefore, efficient algorithms to compute multi-hadron correlation functions are required.
In this work, we use the sequential method \cite{Martinelli:1988} along with stochastic methods \cite{ETM:2008zte} to compute the Dirac propagators. Unfortunately, the former method is not the most efficient way to compute
these multi-hadron correlation functions as many sequential propagators are required to take into account for instance the different momentum combinations and polarizations. There are promising and versatile alternatives to the methods used in this work like distillation \cite{HadronSpectrum:2009krc}, stochastic distillation \cite{Morningstar:2011ka}, and sparsening methods. The latter is employed in \cite{Grebe:2023tfx} to include $N\pi$ and $N\pi\pi$ operators in the variational analysis.

Although the final goal must be to include all the relevant operators in the variational basis, investigations of the variational analysis with the most relevant operators can still provide quite accurate results on the final quantities of interest.
As we discussed in the previous section, ChPT can be used to predict which are the most relevant operators for a given specific channel. In particular, LO-ChPT predicts that $N\pi$ states play an important role in the determination of pseudoscalar and axial-vector form factors from nucleon three-point functions at source-sink separations that are can currently be reached ($\approx 1.5~\mathrm{fm}$).
Motivated by these different studies, we construct a basis with just two operators that have very large overlap with $N$ and $N\pi$ states, respectively. We then compute a $2\times2$ matrix of two-point functions to find the GEVP eigenvalues and eigenvectors. The latter are then used to project the interpolating operators to the GEVP eigenstates $N$ and $N\pi$, and to extract $\langle N|\mathcal{J} | N\rangle $, $\langle N\pi|\mathcal{J} | N\rangle $, by means of eq.~\eqref{gevp_ratio_nn} or \eqref{gevp_ratio_npin}, respectively.

\section{Lattice simulations}\label{sec4}
\subsection{Details of the simulation and construction of operators}
We simulate QCD with $N_f=3$ degenerate quarks using an $\mathcal{O}(a)$-improved Wilson-Clover action. The measurements are performed on 800 gauge configurations 
of the CLS\footnote{See \href{https://wiki-zeuthen.desy.de/CLS/CLS}{https://wiki-zeuthen.desy.de/CLS/CLS} for more information.} ensemble \textit{A653}, with $m_\pi = 420$ MeV, $N_t=2 N_s = 48a$ and lattice spacing $a=0.098$ fm, see \cite{RQCD:2022xux}.
We investigate processes $\langle I'_z; J^{PC} | \mathcal{J}^- | I_z; J^{PC}) \rangle $ 
with $I_z=-\frac{1}{2} = -I'_z$, and $J^{PC}=\frac{1}{2}^{++}$, i.e. lattice irrep $G_1^{(g)}$. 
To this end, we construct two operators $\oo_{3q}$ and $\oo_{5q}$ to have nucleon quantum numbers ($J^{PC}=\frac{1}{2}^{++}$).
The three-quark operator is the standard nucleon operator $\oo_{3q}^\gamma = \epsilon_{abc} \left[u^{\top} C\gamma_5 d \right] q_\gamma$,
with $q=u,d$ for proton or neutron operators,
while the five-quark operator is a nucleon-pion-like operator $\oo_{5q} \sim \oo_{3q} \oo_{\bar{q} q}$ with $\oo_{\bar{q} q} \sim \bar{q} \gamma_5 q$ .
For the analysis, we construct operators in the rest frame ($\pp=\zzero$) and in six equivalent moving frames with discretised unit lattice momentum $\pp=\ei := \pm \frac{2\pi}{L} \hat{n}_i$, where $\hat{n}_i$ is a unit vector along the direction $i=x,y,z$.
The two-particle operator is projected to zero momentum by constructing the composite operator according to the lattice irrep $G_1^+$ of $^2O_h$, the double cover of the cubic group $O_h$, see \cite{Prelovsek:2016iyo, Lang:2016hnn}.
The projection to unit lattice momentum is done by constructing the two-particle operator according to the lattice irrep $G_1$ of the little group $C_{4v}$, see \cite{Gockeler:2012yj}.
Here is a summary of these constructions:
\subparagraph{Rest frame}
\begin{align}
\nonumber
\oo_{5q}^{G_1^+, \uparrow}(\pp=\zzero)
&=
\oo^{\downarrow}_{3q}(-\ex) \oo_{\bar{q}q}(\ex)
-
\oo^{\downarrow}_{3q}(\ex) \oo_{\bar{q}q}(-\ex)
-i
\oo^{\downarrow}_{3q}(-\ey) \oo_{\bar{q}q}(\ey)
~+
\\
\label{lattice_irrep_proj1}
&
+i
\oo^{\downarrow}_{3q}(\ey) \oo_{\bar{q}q}(-\ey)
+
\oo^{\uparrow}_{3q}(-\ez) \oo_{\bar{q}q}(\ez)
-
\oo^{\uparrow}_{3q}(\ez) \oo_{\bar{q}q}(-\ez)
~,
\\
\nonumber
\oo_{5q}^{G_1^+, \downarrow}(\pp=\zzero)
&=
\oo^{\uparrow}_{3q}(-\ex) \oo_{\bar{q}q}(\ex)
-
\oo^{\uparrow}_{3q}(\ex) \oo_{\bar{q}q}(-\ex)
+i
\oo^{\uparrow}_{3q}(-\ey) \oo_{\bar{q}q}(\ey)
~+
\\
\label{lattice_irrep_proj2}
&
-i
\oo^{\uparrow}_{3q}(\ey) \oo_{\bar{q}q}(-\ey)
-
\oo^{\downarrow}_{3q}(-\ez) \oo_{\bar{q}q}(\ez)
+
\oo^{\downarrow}_{3q}(\ez) \oo_{\bar{q}q}(-\ez)
~,
\end{align}
\subparagraph{Moving frame}
\begin{align}
\label{lattice_irrep_proj3}
\oo_{5q, 1}^{G_1, \uparrow / \downarrow}(\pp=\ei)
&= \pm
\oo^{\uparrow / \downarrow}_{3q}(\ei) \oo_{\bar{q}q}(\zzero)~,
\\
\label{lattice_irrep_proj4}
\oo_{5q, 2}^{G_1, \uparrow / \downarrow}(\pp=\ei)
&= \pm
\oo^{\uparrow / \downarrow}_{3q}(\zzero) \oo_{\bar{q}q}(\ei)~.
\end{align}
The superscripts $\uparrow / \downarrow$ refer to the polarization of the spin.
The momentum projection of $3q$- and $\bar{q}q$-operators is performed via Fourier transform 
\begin{equation}
\oo_{3q}(\pp_1) \oo_{\bar{q}q}(\pp_2)
=
\sum_\xx e^{-i \pp_1 \cdot \xx}\oo_{3q}(\xx)
\sum_\yy e^{-i \pp_2 \cdot \yy} \oo_{\bar{q}q}(\yy)~,
\end{equation}
which requires the two operators to be located at different spatial position on the same timeslice. 
While nucleons have isospin $I_N=\frac{1}{2}$, pions have $I_\pi=1$. 
In general, the five-quark interpolator has overlap with isospin states $I=\frac{1}{2} \oplus 1 = \frac{1}{2}, \frac{3}{2}$. 
We use the Clebsch-Gordan coefficients to project the two-hadron operator to isospin $I=\frac{1}{2}$ and $I_z=\pm \frac{1}{2}$, to obtain
\begin{align}
\label{5qp}
\left(I=\frac{1}{2}, I_z=+\frac{1}{2}\right)
\hspace{2cm}
&\oo_{5q}^{p} = +\sqrt{\frac{2}{3}}\oo_{3q}^{p} \oo_{\bar{q}q}^{\pi^0} 
- \frac{1}{\sqrt{3}}\oo_{3q}^{n} \oo_{\bar{q}q}^{\pi^+}~,
\hspace{2cm}
\\
\label{5qn}
\left(I=\frac{1}{2}, I_z=-\frac{1}{2}\right)
\hspace{2cm}
&\oo_{5q}^{n} = -\frac{1}{\sqrt{3}} \oo_{3q}^{n} \oo_{\bar{q}q}^{\pi^0} 
+ \sqrt{\frac{2}{3}} \oo_{3q}^{p} \oo_{\bar{q}q}^{\pi^-}~,
\hspace{2cm}
\end{align}
where the superscripts $p, n, \pi $ are short-hand notations for proton, neutron, and pion isospin quantum numbers. 
The expressions for these standard operators read
\begin{align}
\label{proton_op}
&\oo_{3q}^p(x)^\gamma = \epsilon_{abc} \left[u^{\top}(x)^a_\alpha (C\gamma_5)^{\alpha \beta} d(x)^b_\beta \right] q(x)^c_\gamma~,
\\
\label{neutron_op}
&\oo_{3q}^n(x)^\gamma = \epsilon_{abc} \left[u^{\top}(x)^a_\alpha (C\gamma_5)^{\alpha \beta} d(x)^b_\beta \right] d(x)^c_\gamma~,
\\
\label{piminus_op}
&\oo_{\bar{q}q}^{\pi^-}(y) = \bar{u}(y) \gamma_5 d(y)~,
\\
&\oo_{\bar{q}q}^{\pi^-}(0) = 
\frac{1}{\sqrt{2}} \left[ \bar{u}(y) \gamma_5 u(y) - \bar{d}(y) \gamma_5 d(y)\right]~,
\\
\label{piplus_op}
&\oo_{\bar{q}q}^{\pi^+}(y) = -\bar{d}(y) \gamma_5 u(y)~.
\end{align}
The minus sign in eq.~\eqref{piplus_op} is derived by applying the ladder isospin operator $I^+$ to the operator in eq.~\eqref{piminus_op}.
We combine the isospin projections with the lattice irrep projections in eqs.~\eqref{lattice_irrep_proj1}-\eqref{lattice_irrep_proj4} to construct the final two-hadron operator with the quantum numbers of protons and neutrons.
To construct extended $N$- and $N\pi$-like operators we apply Gaussian smearing to the quark fields \cite{GUSKEN1990361} and APE smoothing to the gauge fields \cite{APE:1987ehd}.  
We use the smearing parameters in \cite{RQCD:2022xux}.

\subsection{Computation of correlation functions}
Using the single- and two-hadron operators in eqs.~\eqref{5qp}-\eqref{neutron_op} we compute the matrix of three-point correlation functions
\begin{equation}\label{gevp_3pts}
	\mathbb{C}_{3pt}(\pp', t; \qq, \tau) =
	\begin{pmatrix}
		\langle \oo_{3q}(\pp', t)~\mathcal{J}(\qq, \tau) ~\bar{\oo}_{3q}(\pp, 0) \rangle  & 
		\langle \oo_{3q}(\pp', t)~\mathcal{J}(\qq, \tau) ~\bar{\oo}_{5q}(\pp, 0) \rangle \\
		\langle \oo_{5q}(\pp', t)~\mathcal{J}(\qq, \tau) ~\bar{\oo}_{3q}(\pp, 0) \rangle  &
		\langle \oo_{5q}(\pp', t)~\mathcal{J}(\qq, \tau) ~\bar{\oo}_{5q}(\pp, 0) \rangle
	\end{pmatrix}~.
\end{equation}
\begin{figure}[t]
	\begin{minipage}{0.5\textwidth}
		\begin{tikzpicture}[scale=0.7]
\filldraw[fill=green!20!white, draw=green!50!black] (0,0.5) circle[x radius=0.5, y radius=1.5];
\filldraw[fill=blue!20!white, draw=blue!50!black] (7,0.5) circle[x radius=0.5, y radius=1.5];
\filldraw[fill=red!20!white, draw=red!50!black] (0,-3.5) circle[x radius=0.5, y radius=1] ;
\draw [thick](0,2) -- (3.5,2);
\draw[<-, thick] (3.5,2) -- (7,2);
\draw[thick](0.5,0.5) -- (3.5,0.5);
\draw [<-, thick] (3.5,0.5) -- (6.5,0.5);
%
%\draw[line width=0.3mm](0.5,0.5) -- (4,0.5);
\draw [<-, thick] (3.5,-2.75) -- (7, -1);
\draw [thick] (0., -4.5) -- (3.5,-2.75);
\draw [line width=0.4mm, red] (1.8,-1.55) -- (2.2,-1.95);
\draw [line width=0.4mm, red] (1.8,-1.95) -- (2.2, -1.55);
%

%\draw  (0,-3.5) circle[x radius=0.5, y radius=1] 
\draw [<-][line width=0.3mm, draw=red!80] (0,-1) arc[x radius=20mm, y radius= 7.5mm, start angle=90, end angle=-90];
%

%\node[text width=6cm, anchor=west, right] at (5,0)
%{In this diagram, what can you say about $\angle F$, $\angle B$ and $\angle E$?};
%
\node [] at (-1.2,1.5) {$u(x)$};
\node[] at (-1.2,0.5) {$u(x)$};
\node[] at (-1.2,-0.5) {$d(x)$};
\node[] at (0.,0.5) {$p$};
\node[] at (2.9,-1.3) {$\bar{d}(z)$};
\node[] at (2.9,-2.) {$u(z)$};
\node[] at (-1.2,-3) {$\bar{u}(y)$};
\node[] at (-1.2,-4) {$d(y)$};
\node[] at (0,-3.5) {$\pi^-$};
\node[] at (8.3,1.5) {$\bar{u}(0)$};
\node[] at (8.3,0.5) {$\bar{u}(0)$};
\node[] at (8.3,-0.5) {$\bar{d}(0)$};
\node[] at (7.,0.5) {$p$};

\node[] at (6,-3.) {\large \textit{A-like}};

\clip (0,2) circle (1cm);

\end{tikzpicture}
	\end{minipage}
	\begin{minipage}{.4\textwidth}
		\begin{tikzpicture}[scale=0.7]
\filldraw[fill=green!20!white, draw=green!50!black] (0,0.5) circle[x radius=0.5, y radius=1.5];
\filldraw[fill=blue!20!white, draw=blue!50!black] (7,0.5) circle[x radius=0.5, y radius=1.5];
\filldraw[fill=red!20!white, draw=red!50!black] (0,-3.5) circle[x radius=0.5, y radius=1] ;
\draw[line width=0.4mm, draw=red!80 ] (0,2.) -- (1.75,2);
\draw[<-, line width=0.4mm, draw=red!80] (1.75,2) -- (3.5,2);

\draw [thick](3.5,2) -- (5.25,2);
\draw [<-, thick] (5.25,2) -- (7,2);
\draw [thick](0.5,0.5) -- (3.5,0.5);
\draw [<-, thick] (3.5,0.5) -- (6.5,0.5);
\draw [<-, thick] (3.5,-2.75) -- (7, -1);
\draw [thick](0., -4.5) -- (3.5,-2.75);
%
%\draw  (0,-3.5) circle[x radius=0.5, y radius=1] 
\draw [<-,line width=0.3mm, draw=red!80!] (0,-1) arc[x radius=20mm, y radius= 7.5mm, start angle=90, end angle=-90];

\draw [line width=0.4mm, red] (3.3,2.2) -- (3.7, 1.8);
\draw [line width=0.4mm, red] (3.3,1.8) -- (3.7, 2.2);

%
%\node[text width=6cm, anchor=west, right] at (5,0)
%{In this diagram, what can you say about $\angle F$, $\angle B$ and $\angle E$?};
%
\node[] at (-1.2,1.5) {$d(x)$};
\node[] at (-1.2,0.5) {$u(x)$};
\node[] at (-1.2,-0.5) {$u(x)$};
\node[] at (0.,0.5) {$p$};
\node[] at (3.,2.6) {$\bar{d}(z)$};
\node[] at (4.1,2.6) {$u(z)$};
\node[] at (-1.2,-3) {$\bar{u}(y)$};
\node[] at (-1.2,-4) {$d(y)$};
\node[] at (0,-3.5) {$\pi^-$};
\node[] at (8.3,1.5) {$\bar{u}(0)$};
\node[] at (8.3,0.5) {$\bar{u}(0)$};
\node[] at (8.3,-0.5) {$\bar{d}(0)$};
\node[] at (7.,0.5) {$p$};

\node[font=\large] at (6,-3.) {\large \textit{B-like}};

\end{tikzpicture}
	\end{minipage}
	\hfill \break
	\\ \hfill \break
	\begin{minipage}{0.5\textwidth}
		\begin{tikzpicture}[scale=0.7]
\filldraw[fill=green!20!white, draw=green!50!black] (0,0.5) circle[x radius=0.5, y radius=1.5];
\filldraw[fill=blue!20!white, draw=blue!50!black] (7,0.5) circle[x radius=0.5, y radius=1.5];
\filldraw[fill=red!20!white, draw=red!50!black] (0,-3.5) circle[x radius=0.5, y radius=1] ;
\draw [line width=0.3mm](0,2) -- (3.5,2);
\draw [<-,line width=0.3mm](3.5,2) -- (7,2);
\draw [line width=0.3mm](0.5,0.5) -- (3.5,0.5);
\draw [<-,line width=0.3mm](3.5,0.5) -- (6.5,0.5);
%

%\draw  (0,-3.5) circle[x radius=0.5, y radius=1] 
\draw [<-,line width=0.4mm, draw=red!80] (0,-1) arc[x radius=20mm, y radius= 7.5mm, start angle=90, end angle=-90];
\draw [<-, line width=0.4mm, draw=red!80] (0.1,-4.5) ..controls (3.6,-2.5) 
and (3.6,-1.2) .. (4.,-1.4);
\draw [thick](4., -1.4) -- (7,-1);
\draw [line width=0.4mm, red] (3.8,-1.2) -- (4.2, -1.6);
\draw [line width=0.4mm, red] (3.8,-1.6) -- (4.2, -1.2);
%
%\node[text width=6cm, anchor=west, right] at (5,0)
%{In this diagram, what can you say about $\angle F$, $\angle B$ and $\angle E$?};
%

\node[] at (-1.2,1.5) {$d(x)$};
\node[] at (-1.2,0.5) {$u(x)$};
\node[] at (-1.2,-0.5) {$u(x)$};
\node[] at (0.,0.5) {$p$};
\node[] at (3.4,-0.9) {$\bar{d}(z)$};
\node[] at (4.5,-0.9) {$u(z)$};
\node[] at (-1.2,-3) {$\bar{u}(y)$};
\node[] at (-1.2,-4) {$d(y)$};
\node[] at (0,-3.5) {$\pi^-$};
\node[] at (8.3,1.5) {$\bar{d}(0)$};
\node[] at (8.3,0.5) {$\bar{u}(0)$};
\node[] at (8.3,-0.5) {$\bar{u}(0)$};
\node[] at (7.,0.5) {$p$};
\node[font=\large] at (6,-3.) {\large \textit{C-like}};

\end{tikzpicture}
	\end{minipage}
	\begin{minipage}{0.4\textwidth}
		\begin{tikzpicture}[scale=0.7]
\filldraw[fill=green!20!white, draw=green!50!black] (0,0.5) circle[x radius=0.5, y radius=1.5];
\filldraw[fill=blue!20!white, draw=blue!50!black] (7,0.5) circle[x radius=0.5, y radius=1.5];
\filldraw[fill=red!20!white, draw=red!50!black] (0,-2.8) circle[x radius=0.5, y radius=1] ;
\draw  [line width=0.3mm](0,2) -- (3.5,2);
\draw  [<-,line width=0.3mm](3.5,2) -- (7,2);
\draw [line width=0.3mm] (0.5,0.5) -- (4.,0.5);
\draw [<-, line width=0.3mm](3.5,0.5) -- (6.5,0.5);

\draw [line width=0.3mm](0,-1) -- (3.5, -1);
\draw [<-, line width=0.3mm](3.5,-1) -- (7, -1);

\draw (0,-1.8) arc[x radius=35mm, y radius= 10mm, start angle=90, end angle=-90][line width=0.4mm, red];
\draw [ line width=0.4mm, red] (3.3,-2.6) -- (3.7, -3.);
\draw [ line width=0.4mm, red] (3.3,-3.) -- (3.7, -2.6);
%
%\node[text width=6cm, anchor=west, right] at (5,0)
%{In this diagram, what can you say about $\angle F$, $\angle B$ and $\angle E$?};
%

\node[] at (-1.2,1.5) {$u(x)$};
\node[] at (-1.2,0.5) {$u(x)$};
\node[] at (-1.2,-0.5) {$d(x)$};
\node[] at (0.,0.5) {$p$};
\node[] at (4.5,-3.2) {$\bar{d}(z)$};
\node[] at (4.5,-2.4) {$u(z)$};
\node[] at (-1.2,-2.4) {$\bar{u}(y)$};
\node[] at (-1.2,-3.2) {$d(y)$};
\node[] at (0,-2.8) {$\pi^-$};
\node[] at (8.3,1.5) {$\bar{u}(0)$};
\node[] at (8.3,0.5) {$\bar{u}(0)$};
\node[] at (8.3,-0.5) {$\bar{d}(0)$};
\node[] at (7.,0.5) {$p$};

\node[font=\large] at (6.4 ,-2.5) {\large \textit{D-like}};

\end{tikzpicture}
	\end{minipage}
	\hfill \break
	\caption{\\
		This schematic plot represents the topologies $A$, $B$, $C$ and $D$ of the Wick contractions 
		in the correlation functions $\langle \oo_{5q}^n(\pp', t)~\mathcal{J}^-(\qq, \tau) ~\bar{\oo}^p_{3q}(\pp, 0) \rangle$.
		The red lines are all-to-all propagators, two for each diagram, and the black lines are point-to-all propagators.
		The red cross is the current insertion.}
	\label{figure:diagrams_protonJminus_to_proton_piminus}
\end{figure}
We neglect the lower diagonal entry $\langle \oo_{5q}(\pp', t)~\mathcal{J}(\qq, \tau) ~\bar{\oo}_{5q}(\pp, 0) \rangle$
because it corresponds to $N\pi \to N\pi$ terms, which are expected to be volume suppressed in the standard three-point functions
$\langle \oo_{3q}(\pp', t)~\mathcal{J}(\qq, \tau) ~\bar{\oo}_{3q}(\pp, 0) \rangle$,
and therefore their contribution will be much smaller.
Also, it is numerically more demanding to compute these correlation functions with our methods as it involves several Wick contractions and all-to-all propagators.
The off-diagonal entries in eq.~\eqref{gevp_3pts} correspond to the current-enhanced terms (when allowed) in the standard three-point functions,
i.e., $N\to N\pi$ and $N\pi \to N$ states.
The Wick contractions of $\langle \oo_{5q}(\pp', t)~\mathcal{J}(\qq, \tau) ~\bar{\oo}_{3q}(\pp, 0) \rangle$ were first discussed in \cite{Barca:2021iak},
and they can be grouped in 4 diagrams: A, B, C, and D, see Fig.~\ref{figure:diagrams_protonJminus_to_proton_piminus}.
The diagrams contain 2 all-to-all propagators and we adopt the sequential method \cite{Martinelli:1988} with 2 sequential propagators to compute the diagrams A, B, and C.
The diagram D is much cheaper to compute because we use the point-to-all method for the nucleon part and the \textit{one-end-trick} \cite{ETM:2008zte}
with 12 stochastic vectors for the pion-to-current term.
In particular, our lattice simulations show that for the diagram D
\begin{equation}
	\langle \oo_{5q}(\pp', t)~\mathcal{J}(\qq, \tau) ~\bar{\oo}_{3q}(\pp, 0) \rangle_D
	\approx 
	\langle \oo_{3q}(\pp'_N, t)~\bar{\oo}_{3q}(\pp, 0) \rangle
	~
	\langle \oo_{\bar{q}q}(\pp'_\pi, t)~\mathcal{J}(\qq, \tau)\rangle~.
\end{equation}
This factorisation, together with the observation that the diagram D has the largest signal,
explains which channels have the large $N\pi$ contamination in standard nucleon three-point functions.
In the case of the axial-vector current $\mathcal{J}=\mathcal{A}_\mu$, the diagram D reads
\begin{align}\label{c3pt_diagramD_fact}
	\langle \oo_{5q}(\pp', t)~\mathcal{A}_\mu(\qq, \tau) ~\bar{\oo}_{3q}(\pp, 0) \rangle_D
	& \propto  \delta_{\pp', \pp} ~ \delta_{\pp'_\pi, \qq} ~ |\langle \Omega | \oo_{3q}|N\rangle |^2 i f_\pi q_\mu ~e^{-E_N t}  e^{-E_\pi (t-\tau)}
\end{align}
which is non-zero only when $\pp'_N=\pp$, $\pp'_\pi = \qq$ and $q_\mu \neq 0$.
This explains the large ($N\pi$) excited state contamination in the $\mathcal{A}_4$ channel,
and the milder excited state contamination in the channel with $\mathcal{A}_i$ and $q_i = 0$,
where the momentum of the current is zero along the direction where the spin is aligned and the current component,
see Fig.~4 of \cite{RQCD:2019jai}. Similar arguments hold with a pseudoscalar current.
\begin{figure}[t]
	\includegraphics[width=\textwidth]{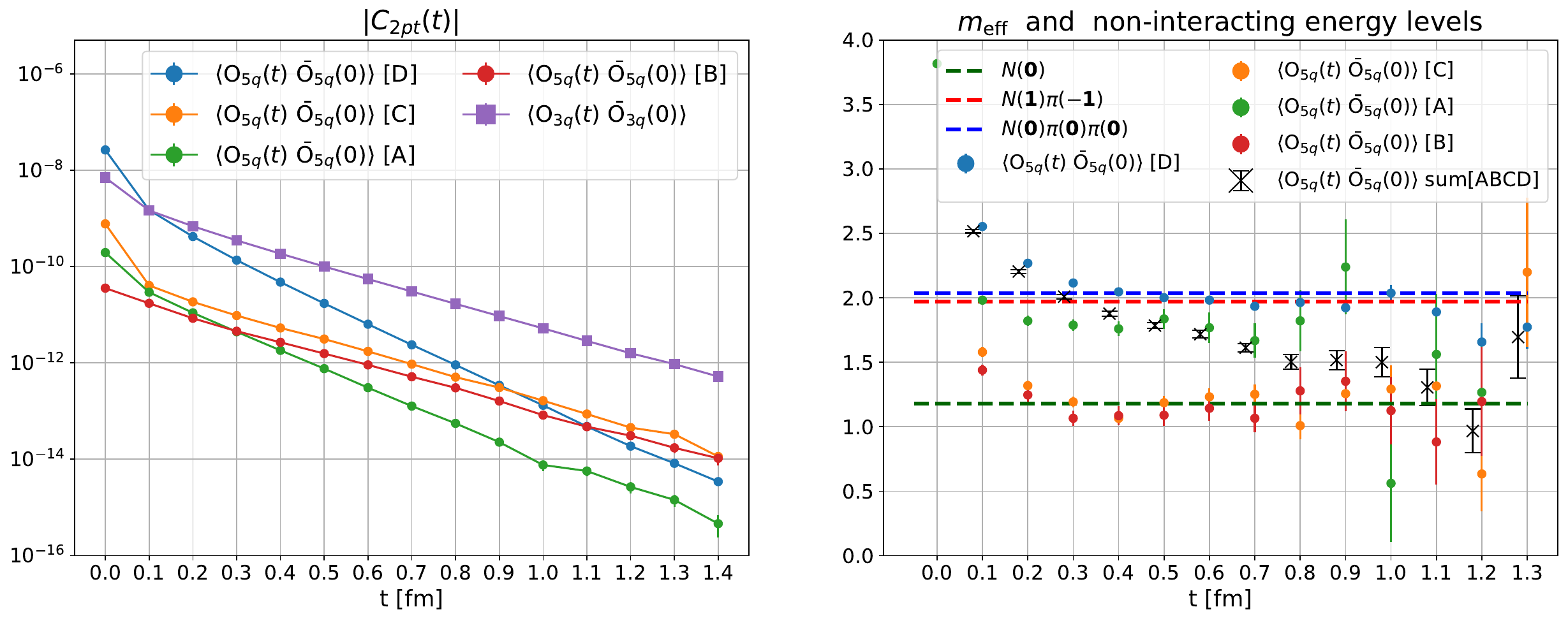}
	\caption{\\ Two-point correlation functions on the left and respective effective masses on the right.
		The diagram B and C, which consists of $3$-quark propagators from time $0$ to $t$ are decaying immediately
		with the nucleon energy, while diagrams A and D, which consists of $4$-quark and $1$-antiquark propagators 
		from $0$ to $t$ are decaying with an interacting energy close to the lowest non-interacting $N\pi$ (and $N\pi\pi$).
	}
	\label{figure:comparison_2pts_A653}
\end{figure}
In order to use the variational method, we compute the $2\times2$ GEVP matrix
\begin{equation}\label{gevp_2pts}
\mathbb{C}_{2pt}(\pp, t) =
\begin{pmatrix}
	\langle \oo_{3q}(\pp, t) ~\bar{\oo}_{3q}(\pp, 0) \rangle  & \langle \oo_{3q}(\pp, t) ~\bar{\oo}_{5q}(\pp, 0) \rangle \\
	\langle \oo_{5q}(\pp, t) ~\bar{\oo}_{3q}(\pp, 0) \rangle  & \langle \oo_{5q}(\pp, t) ~\bar{\oo}_{5q}(\pp, 0) \rangle \\
\end{pmatrix}~,
\end{equation}
see discussion in sec.~\eqref{sec3}.
The correlation functions $\langle \oo_{3q}(\pp, t) ~\bar{\oo}_{5q}(\pp, 0) \rangle$ are constructed from $\langle \oo_{3q}(\pp', t)~\mathcal{J}(\qq, \tau) ~\bar{\oo}_{3q}(\pp, 0) \rangle$ with $\mathcal{J}=\mathcal{P}$, $\tau=0$, and at different source-sink separations $t$.
Similarly, the ($N\pi$-like) two-point correlation functions $\langle \oo_{5q}(\pp, t) ~\bar{\oo}_{5q}(\pp, 0) \rangle$ are constructed from the three-point functions $\langle \oo_{5q}(\pp', t)~\mathcal{J}(\qq, \tau) ~\bar{\oo}_{3q}(\pp, 0) \rangle$; to visualise this construction see Fig.~\ref{figure:diagrams_protonJminus_to_proton_piminus} 
with the current $\mathcal{J}=\mathcal{P}$ at the source ($\tau=0$).
The sequential method, which is used in this pilot study, is not the most efficient approach and in the future other methods 
like distillation \cite{HadronSpectrum:2009krc} will be considered, 
especially when $N\pi\pi$ operators will be included in the basis.
In Fig.~\ref{figure:comparison_2pts_A653}, there is a comparison plot between the diagrams A, B, C, and D 
that make up the $N\pi$-like two-point functions and the standard nucleon-like two-point functions.
On the right plot, the effective energies of the diagrams and their sum are compared against $N$, $N\pi$ and $N\pi\pi$ energies in the non-interacting limit. The operators are projected to total momentum zero in these plots.
While the diagrams A and D decay with some interacting $N\pi$ energy which is very close to the non-interacting $N\pi$ energy (or $N\pi\pi$ energy 
as they are degenerate in this finite volume), the diagrams B and C decay quite fast with the nucleon energy. 
This behaviour is expected, see e.g. Fig.~7 of \cite{Liu:2020okp}, 
as the diagrams A and D are constructed from 5 propagators travelling from time $0$ to time $t$, 
while the diagrams B and C contain just 3 propagators from time $0$ to time $t$ and 2 propagators from time $t$ to time $t$, see Fig.~\ref{figure:diagrams_protonJminus_to_proton_piminus}.
Eventually, the sum of all the diagrams that compose the $N\pi$-like two-point function will decay at large time with the nucleon energy as the $N\pi$ operators are constructed to have the quantum numbers of the nucleon.

\subsection{GEVP results}
\begin{figure}[ht!]
	\centering
	\includegraphics[width=\textwidth]{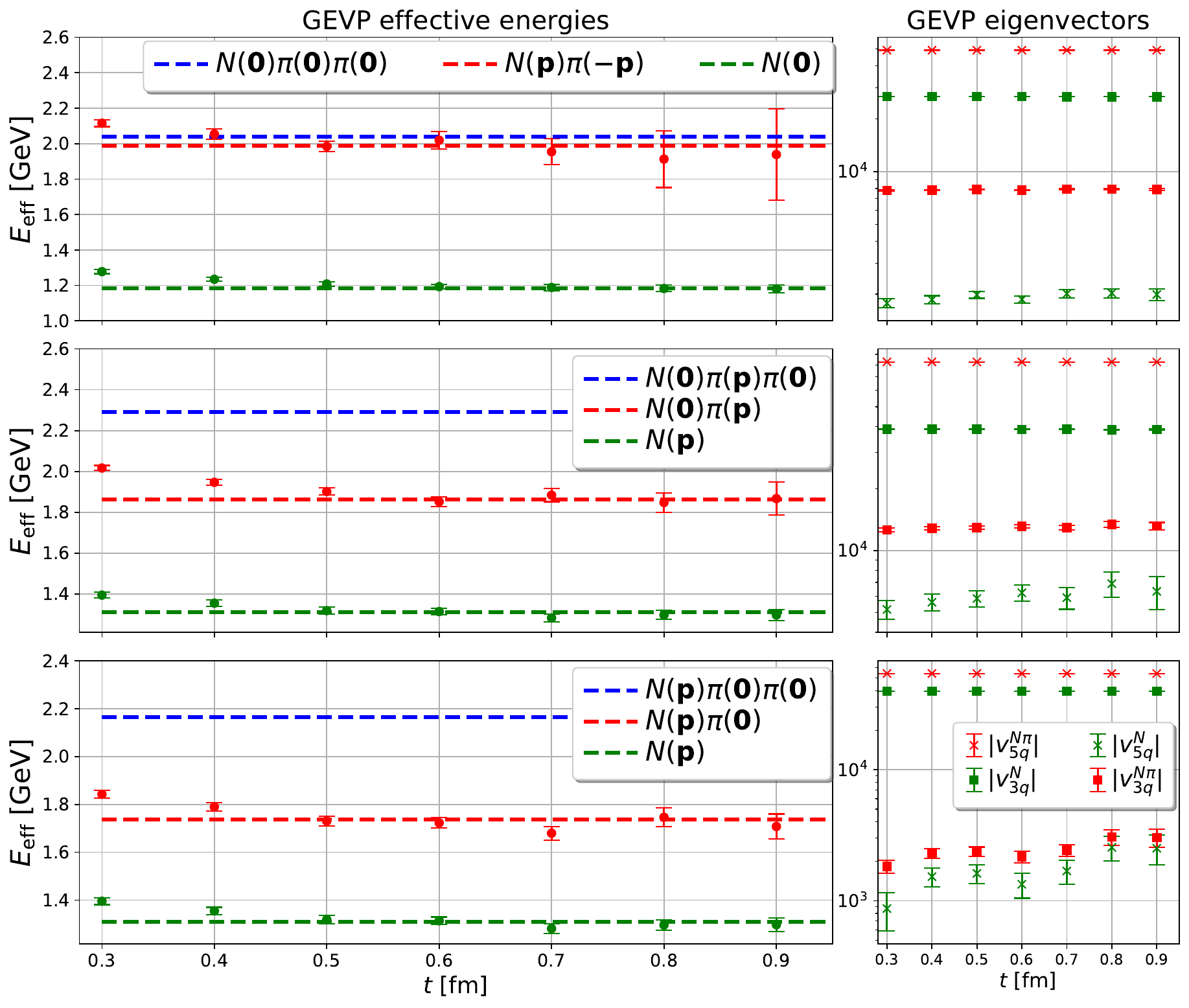}
	\caption{Effective energies of GEVP eigenvalues (green and red data points) on the left, compared to $N\pi$ and $N\pi\pi$ non-interacting energies, as well as nucleon energies extracted from nucleon two-point functions. The pion energies are extracted from pion two-point functions.
		(top) Energies in the rest frame with total momentum zero.
		(middle) GEVP energies in the moving frame using the $5q$ operator in eq.~\eqref{lattice_irrep_proj4}, where the nucleon operator is projected to zero momentum.
		(bottom) GEVP energies in the moving frame using the $5q$ operator in eq.~\eqref{lattice_irrep_proj3}, where the pion operator is projected to zero momentum.
		The relative momentum is $\mathbf{p}=\frac{2\pi}{L} \hat{n}$, with $\hat{n}$ a unit vector along the 3 spatial directions.
		On the right, GEVP eigenvectors normalised w.r.t. $\mathbb{C}_{2pt}(t_0)$ as in eq.~\eqref{normalised_eigenvectors}.
	}
	\label{figure:gevp_results}
\end{figure}

We solve the GEVP in eq.~\eqref{gevp} using the matrix in eq.~\eqref{gevp_2pts} to determine the matrix of eigenvectors $V(\pp, t;t_0)$ and eigenvalues $\Lambda(\pp, t; t_0)$.
In Fig.~\ref{figure:gevp_results}, the GEVP results at $t_0=0.2~\mathrm{fm}$ are plotted for $\pp=\zzero$ and $\pp = \ei$ ($|\pp| \approx 520~\mathrm{MeV}$). 
For the kinematic frame with momentum $\pp$, we use both the $5q$-operators in eqs.~\eqref{lattice_irrep_proj3}-\eqref{lattice_irrep_proj4}.
On the left plots, the effective energies of the 2 GEVP eigenvalues are compared to the $N$, $N\pi$ and $N\pi\pi$ energies in the non-interacting limit.
The nucleon and pion energies are extracted at large time from nucleon and pion two-point functions.
In the rest frame with zero total momentum, the energy of the first eigenvalue matches the nucleon energy. 
The second eigenvalue has an energy very close to that of the non-interacting $N\pi$ P-wave state and the 
$N\pi\pi$ S-wave state, which are (quasi-)degenerate in this ensemble. 
This comparison between $N\pi$ and $N\pi\pi$ states in the non-interacting limit 
varies with the pion mass and the physical volume, see e.g. Fig.~2 in \cite{Green:2018vxw}.
In the moving frames with total momentum $\pp$, the first eigenvalue is again consistent with the nucleon state and the second eigenvalue is more clearly a $N\pi$ state, due to the non-degeneracy of the $N\pi$ and $N\pi\pi$ states in the non-interacting limit.
We thus refer to the two GEVP eigenvalues as $\lambda^N$ and $\lambda^{N\pi}$.
The GEVP results for the moving frames in Fig.~\ref{figure:gevp_results} are averaged over equivalent momenta ($\pp=\hat{e}_x, \hat{e}_y, \hat{e}_z, -\hat{e}_x, -\hat{e}_y, -\hat{e}_z$).
We have to stress that the ensemble is at the $SU(3)$ symmetric point and this higher symmetries imply more degeneracy with other physical states, 
e.g. $\Lambda K$ are degenerate to $N\pi$. This degeneracy could in principle affect the GEVP eigensolutions and must be taken into account to reduce 
systematic effects.

On the right plots, the components of the normalised GEVP eigenvectors are presented.
The eigenvectors are normalised according to eq.~\eqref{normalised_eigenvectors} and
we refer to them as $\tilde{v}^N$ and $\tilde{v}^{N\pi}$. In particular, the components are
$\tilde{v}^N=(\tilde{v}^N_{3q}, \tilde{v}^N_{5q})$ and $\tilde{v}^{N\pi}=(\tilde{v}^{N\pi}_{3q}, \tilde{v}^{N\pi}_{5q})$. 
%As expected from arguments of volume suppression in the phase space, we find the hierarchy $|\tilde{v}^N_{3q}|> |\tilde{v}^{N\pi}_{3q}|$, and $|\tilde{v}^{N\pi}_{5q}| > |\tilde{v}^{N}_{5q}|$.
%\renewcommand{\arraystretch}{1.5}

\begin{comment}
\begin{table}[ht!]
	\renewcommand{\arraystretch}{1.5}
	\centering
	\begin{tabular}{|c|c|c|c|c|c|}
		\hline
		GEVP basis & GEVP state & $|v_{3q}^{N}|$ & $|v_{3q}^{N\pi}|$ & $|v_{5q}^{N}|$ & $|v_{5q}^{N\pi}|$ \\
		\hline
		$\oo^{G_1^+}_{3q}(\pp=\zzero)$, $\oo^{G_1^+}_{5q}(\pp=\zzero)$ & $N(\zzero), N(\pp)\pi(-\pp)$ & 0.9973(3)  &  0.1587(11) & 0.0735(39) & 0.9873(2) \\
		\hline
		$\oo^{G_1}_{3q}(\pp=\ei)$, $\oo^{G_1}_{5q, 2}(\pp=\ei)$  & $N(\pp), N(\zzero)\pi(\pp)$ & 0.9870(11)    & 0.1581(15)  & 0.1604(70)  & 0.9874(2) \\
		\hline
		$\oo^{G_1}_{3q}(\pp=\ei)$, $\oo^{G_1}_{5q, 1}(\pp=\ei)$  & $N(\pp), N(\pp)\pi(\zzero)$ & 0.9993(2)   & 0.0412(23)   & 0.0379(43) & 0.9991(1) \\
		\hline
	\end{tabular}
	\caption{\\Components of GEVP eigenvectors relative to $3q$- and $5q$-operators with total momentum $\pp=\zzero$, $\ei(= \frac{2\pi}{L}\hat{n}_i)$. \textbf{TO-DO: Replace $v_i^\alpha$ with $Z_i^\alpha$ and compare with ChPT.}}
\end{table}
\end{comment}

\subsection{GEVP ratios $\langle N| \mathcal{J}| N \rangle $}
Using the eigenvectors $v^N(\pp)$ in the plateau region, 
i.e. $0.5~\mathrm{fm} \lesssim t \lesssim 0.9~\mathrm{fm}$ in Fig.~\ref{figure:gevp_results}, 
we construct the GEVP-projected interpolating operator
\begin{equation}
\oo^N(\pp) = \sum_i v_i^N(\pp) \oo_i(\pp)
\hspace{6em}
\bar{\oo}^N(\pp) = \sum_i v_i^N(\pp) \bar{\oo}_i(\pp)
\end{equation}
where we emphasize that the eigenvectors are fairly constant with $t$, 
and use eqs.~\eqref{gevp_corr2pt}-\eqref{gevp_corr3pt} to construct the GEVP-improved two- and three-point correlation functions
\begin{align}
\label{gevp_projected_n2pt}
C_{2pt}(\pp, t)^N &= \langle \oo^N(\pp, t) ~ \bar{\oo}^N(\pp, 0) \rangle~,
\\
C^{k, \mathcal{J}}_{3pt}(\pp', t; \qq, \tau)^{N N} &= \langle \oo^N_k(\pp', t) ~ \mathcal{J}(\qq, \tau) ~ \bar{\oo}_k^{N}(\pp) \rangle~.
\end{align}
The subscript $k$ refers to the polarization of the interpolating operators.
After the GEVP-projection, one can check that the two-point functions decay with an energy close to the nucleon energy and 
that the three-point functions decay with $\tau$ with the source-sink energy difference $E_N - E'_N$.
It is important to emphasize that the inclusion of $N\pi$ operators in the variational basis did not improve much the plateau
of the nucleon two-point functions, but it makes a substantial change in the nucleon three-point functions. 

\paragraph{Forward limit} \hfill \break 
The GEVP-improved correlation functions are used to construct the GEVP ratios in eq.~\eqref{gevp_ratio_nn}.
In the forward limit, we set $\pp'=\pp$ and $\qq = \zzero$, and investigate both $\pp = \zzero$ and $\pp=\hat{e}_i$.
We expect from LO-ChPT that only the channels $\mathcal{J}=\mathcal{A}_4, \mathcal{P}$ with $\pp=\pp'=\hat{e}_i$ 
have current-enhanced $N\pi$ contamination, while in the other channels, namely $\mathcal{A}_4$ and $\mathcal{P}$ with $\pp=\zzero$
and $\mathcal{A}_i$ with $\pp=\zzero, \hat{e}_i$ the excited state contamination is overall much smaller.
This expectation is also confirmed by our numerical study, where in \cite{Barca:2022uhi} we report the channels affected by large ESC.
The inclusion in our basis of $N\pi$-like interpolating operators with unit relative momenta has little effect in the determination of $g_A$
with $\mathcal{J}=\mathcal{A}_i$ and with $\pp=\zzero, \hat{e}_i$.
As discussed in the previous section, this is mainly due to the diagram D which is not contributing to these channel when $q_i=0$, 
see eq.~\eqref{c3pt_diagramD_fact}, but the other diagrams have a non-zero but rather small amplitude.

\paragraph{Off-forward limit} \hfill \break 
In the off-forward limit, we expect from LO-ChPT a large $N\pi$ contamination in the channels with $\mathcal{J}=\mathcal{A}_\mu, \mathcal{P}$
and with $q_i\neq0$ where $i$ is the direction of the polarization.
Our variational analysis with $N$- and $N\pi$-like operators confirm again these predictions and it shows that the $N\pi$ operators included in this
work have little effect in the off-forward limit channel $\mathcal{A}_i$ with $q_i=0$, e.g. $\mathcal{A}_{i \neq 3}$, $\mathbb{P}^+_{i\neq3}$ and $\qq = \frac{2\pi}{L}(0,0,1)$ like in Fig.~4 of \cite{RQCD:2019jai}, where the excited state contamination is much smaller.
This can be seen again from the behaviour of the diagram D in eq.~\eqref{c3pt_diagramD_fact}.
A comparison of standard and GEVP-improved ratios in the off-forward limit is presented in \cite{Barca:2022uhi}.
Full results in the forward and off-forward limit will appear in a separate publication.

\begin{comment}
We average the correlation functions over all equivalent momenta combinations.
The pseudoscalar nucleon form factor is extracted with a fit ansatz of the form
\begin{equation}
	R = G_P - A e^{-\Delta E (t-\tau)}
\end{equation}

where the second term fits the remaining excited state contamination at the sink ($N\pi$ with higher relative momenta, $N\pi\pi$, ...)
The pseudoscalar form factor $G_P$, together with the other nucleon form factors extracted from the other channels, 
satisfy the PCAC and PPD relations up to modest $\mathcal{O}(a)$-effects, see \cite{Barca:2022uhi}.
For smaller lattice artifacts, one should $\mathcal{O}(a)$-improve all the entries in the GEVP matrix of three-point functions in eq.~\eqref{gevp_3pts}.
\end{comment}

\subsection{GEVP ratios for the extraction of $\langle N\pi(\pp')| \mathcal{J}(\qq)| N (\pp)\rangle $}
\begin{figure}[t]
	\includegraphics[width=\textwidth]{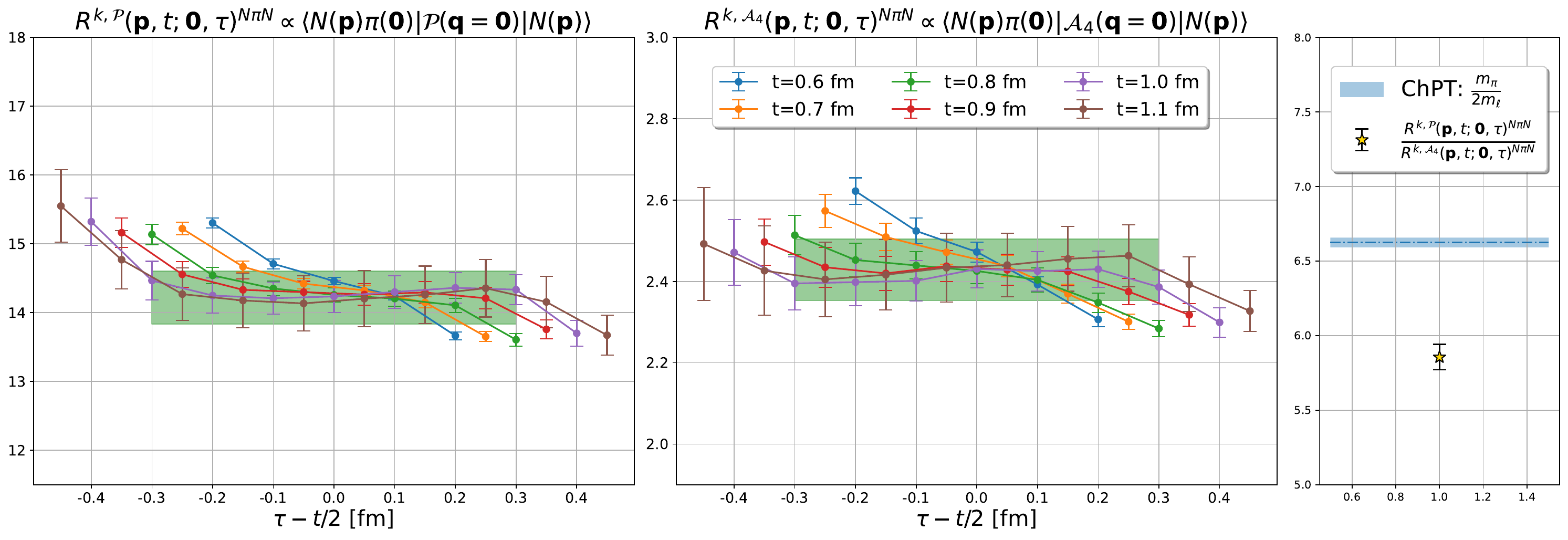}
	\caption{\\ Renormalised GEVP-ratios in eq.~\eqref{gevp_ratio_npin} at $\qq=\zzero$, $\pp=\hat{e}_i$
		using the pseudoscalar (left) and temporal axial-vector (middle) current.
		The GEVP ratios for $\langle N\pi |\mathcal{J} | N \rangle$ in these plots result 
		much flatter than the GEVP ratios for the extraction of $\langle N | \mathcal{J} | N \rangle$, see \cite{Barca:2022uhi}.
		A constant fit using only the largest source-sink separations give for these ratios:
		$R^{i, \mathcal{P}} = 14.22(39)$ and  $R^{i, \mathcal{A}_4} =2.43(8)$.
		On the right plot, there is a comparison between the double ratio, which is equivalent to the ratio of transition matrix elements,
		and its ChPT prediction.}
	\label{figure:gevp_ratio_n2npi}
\end{figure}

We compute for the first time the transition matrix elements $\langle N\pi(\pp) | \mathcal{J}(\zzero) | N(\pp) \rangle$ by constructing the GEVP-improved operators
\begin{align}
\oo^N(\pp) = \sum_{i=3q, 5q} v_i^N(\pp) \oo_i(\pp)
\hspace{6em}
\bar{\oo}^{N\pi}(\pp) = \sum_{i=3q, 5q} v_i^{N\pi}(\pp) \bar{\oo}_i(\pp)
\end{align}
and by computing the GEVP-improved correlator in eq.~\eqref{gevp_projected_n2pt} together with
\begin{align}
C_{2pt}(\pp, t)^{N\pi} &= \langle \oo^{N\pi}(\pp, t) ~ \bar{\oo}^{N\pi}(\pp, 0) \rangle~,
\\
\label{gevp_3pt_n2npi}
C^{k, \mathcal{J}}_{3pt}(\pp, t; \zzero, \tau)^{N\pi N} &= \langle \oo_k^{N\pi}(\pp', t) ~ \mathcal{J}(\qq, \tau) ~ \bar{\oo}_{k}^{N}(\pp) \rangle~.
\end{align}
Then we construct the GEVP-ratio in eq.~\eqref{gevp_ratio_npin} and 
in Fig.~\eqref{figure:gevp_ratio_n2npi} we show the renormalised GEVP-ratios with $\mathcal{J}=\mathcal{A}_4, \mathcal{P}$ for different source-sink separations and with $\pp'=\pp = \hat{e}_k$.
The resulting GEVP ratios are much flatter than the ones for the extraction of $\langle N | \mathcal{J} | N \rangle $,
which seems very promising for future calculations. 
These lattice results can be compared against ChPT, which predicts that
\begin{equation}
\frac{\langle N(\pp)\pi(\zzero) | \mathcal{P}(\zzero) | N(\pp) \rangle}{\langle N(\pp)\pi(\zzero) | \mathcal{A}_4(\zzero) | N(\pp) \rangle}
=
\frac{m_\pi}{2m_\ell}~.
\end{equation}
This relation should hold up to $\mathcal{O}(a)$-effects as the intermediate currents are not $\mathcal{O}(a)$-improved.
The ratio of matrix elements is equivalent to the double ratio 
$\frac{R^{k, \mathcal{P}}(\pp, t; \zzero, \tau)^{N\pi N} }{R^{k, \mathcal{A}_4}(\pp, t; \zzero, \tau)^{N\pi N} } 
= 
\frac{C^{k, \mathcal{P}}_{3pt}(\pp, t; \zzero, \tau)^{N\pi N}}{C^{k, \mathcal{A}_4}_{3pt}(\pp, t; \zzero, \tau)^{N\pi N}}$, 
and the two ratios are plotted on the right panel in Fig.~\ref{figure:gevp_ratio_n2npi}. 
It is reassuring to see that they are not completely in disagreement with ChPT as $\mathcal{O}(a)$-effects and $SU(3)$ symmetry effects must be still taken into account. These transition matrix elements are at a momentum transfer $Q^2=m_\pi^2\approx 0.175 ~\mathrm{GeV}^2$.
Results for the matrix elements with $\qq \neq \zzero$, which are of paramount importance for the analysis of the neutrino oscillation data and more, will be presented in a separate publication.

\section{Conclusions}\label{sec5}
Elastic nucleon form factors $N \to N$, as well as transition form factors for $N\to \Delta$, $N \to N^*$ and $N\to N\pi$ 
are of paramount importance for analysis of neutrino-nucleus scattering data in the resonance region production.
In this study, we used a variational analysis to explicitly remove the $N\to N\pi$ matrix elements from the standard nucleon-like three-point functions.
As a by-product, we present some preliminary results for the transition matrix elements $N\to N\pi$ at vanishing spatial momentum transfer 
for the isospin transition $1/2 \to 1/2$. Results for the elastic nucleon form factors were presented in \cite{Barca:2022uhi}.
We find that with the GEVP analysis the ratios tend to be surprisingly flat at source-sink separations $\gtrsim 0.8~\mathrm{fm}$,
which allowed a simple constant fit to the data at the largest available source-sink separation.
The results suffer from $\mathcal{O}(a)$ effects as the currents are not $\mathcal{O}(a)$-improved.
The ratio of $\mathcal{P}$ and $\mathcal{A_4}$ matrix elements is found to be within $\sim 10\%$ of the ChPT prediction.
Other results at non-vanishing momentum transfer and with other currents will be presented in a separate publication.
Our lattice simulations are performed on a single $N_f=3$ ensemble with lattice spacing 
$a= 0.1~\mathrm{fm}$, $m_\pi=420~\mathrm{MeV}$ and $T=2L= 4.8~\mathrm{fm}$.
The $SU(3)$ symmetry introduces another small complication as the $N\pi$ states are degenerate to $\Lambda K$ states.
Although we expect this to not have any large effect in the three-point functions, they might have some few percent effect
in the nucleon two-point functions and therefore they can introduce some small systematic effects in the GEVP eigensolutions.
This work represents only the first step towards the calculation of transition form factors with a more complete variational analysis,
where single-hadron and multi-hadron operators must be included in a Lüscher formalism to reconstruct the spectra.
The results confirm the ChPT prediction of very large $N\pi$ states contamination in some nucleon-like three-point functions 
and we emphasize that one particular lattice diagram (D) is the most important one.

\paragraph{Acknowledgements}
L. B. would like to thank the staff members of the lattice group at the Lawrence Berkeley National Laboratory,
namely K.-F. Liu, W. Bigeng, R. Brice\~no, A. Walker-Loud, and D. Pefkou,
for their warm hospitality and for the interesting discussions.
L. B. has benefited from discussions with O. B\"ar, R. Brice\~no, J. Green, R. Gupta, L. Leskovec, and K.-F. Liu.
The work is supported by the German Research Foundation (DFG) research unit FOR5269
"Future methods for studying confined gluons in QCD".
Simulations were performed on the QPACE~3 computer of SFB/TR\nobreakdash-55, using an
adapted version of the {\sc Chroma}~\cite{Edwards:2004sx} software package.

\bibliography{bibliography}
\bibliographystyle{unsrt}

\end{document}